\documentclass[a4paper,12pt]{article}
\usepackage{amssymb}
\usepackage{amsmath}
\usepackage{ascmac}
\usepackage{comment}
\usepackage{wrapfig}
\newtheorem{thm}{Theorem}[section]
\newtheorem{prop}[thm]{Proposition}
\newtheorem{lem}[thm]{Lemma}

\newtheorem{rem}[thm]{Remark}

\def\C{{\mathbb C}}
\def\P{{\mathbb P}}
\def\Z{{\mathbb Z}}
\def\dfrac#1#2{{\displaystyle\frac{#1}{#2}}}

\def\o#1{{\overline{#1}}}
\def\u#1{{\underline{#1}}}
\def\ds{\displaystyle}
\def\prf{\noindent{\bf Proof.} }

\makeatletter
\@addtoreset{equation}{section}

\makeatother
\begin{document}

\begin{center}
{\Large {\bf Study of $q$-Garnier system by Pad\'e method}}

\vspace{10mm}
{\large By}

\vspace{5mm}
{\large Hidehito Nagao$^1$ and Yasuhiko Yamada$^2$}

(Akashi College$^1$ and Kobe University$^2$, Japan)
\end{center}

\noindent
{\bf Abstract.}
We give a simple form of the evolution equation and a scalar Lax pair for the $q$-Garnier system. Some reductions to the $q$-Painlev\'e equations and the autonomous case as a generalized QRT system are discussed.
Using two kinds of Pad\'e problems on differential grid and $q$-grid, we derive some special solutions of the $q$-Garnier system in terms of the  $q$-Appell Lauricella function and the generalized $q$-hypergeometric function. 



%\maketitle
%\tableofcontents
\renewcommand\baselinestretch{1.2}

\section{Introduction}

The Garnier system \cite{Garnier1912, IKSY91} is known as an important extension of the Painlev\'e equations  to multi-variables.
Its $q$-difference analog, the $q$-Garnier system,  was formulated by H.Sakai in \cite{Sakai05-1}.

There exists a simple method to study the Painlev\'e/Garnier equations using Pad\'e approximation \cite{Yamada09}.
In this method, one can obtain the evolution equation, the Lax pair and some special solutions simultaneously, starting from a suitable Pad\'e approximation (or interpolation) problem.
This method has been applied \cite{Ikawa13, Nagao15-1, Nagao15-2, Nagao16, NTY13, Yamada14} to various cases of discrete Painlev\'e equations \cite{KNY15,Sakai01}. 

Our aim is to study the $q$-Garnier system applying the Pad\'e method. We study both the usual (i.e. differential) Pad\'e approximation and the Pad\'e interpolation on $q$-grid, and obtain two kinds of special solutions written in terms of the $q$-Appell Lauricella function and the generalized $q$-hypergeometric function.

In Section \ref{subsec:LG}, we introduce a scalar Lax pair and derive the $q$-Garnier equation as the necessary condition for the compatibility. In Section \ref{subsec:Sakai}, the correspondence to the Sakai's matrix Lax form is considered. In Section \ref{subsec:sufficiency}, the sufficiency for the compatibility is proved. In Section \ref{subsec:root}, we rewrite the $q$-Garnier system into more explicit (but nonbirational) form. In Section \ref{subsec:dege}, we discuss the reductions to the $q$-Painlev\'e equations of types $D_5^{(1)}$, $E_6^{(1)}$ and $E_7^{(1)}$. The $E_7^{(1)}$ case is new and $D_5^{(1)}$, $E_6^{(1)}$ cases are known \cite{Sakai05-1, Sakai06}. 

In Section \ref{sec:QRT}, we formulate a hyper-elliptic generalization of the QRT system \cite{QRT89,Tsuda04}. Then the generalized QRT system is identified as the autonomous limit of the $q$-Garnier system. 

In Section \ref{sec:padea}, we study certain Pad\'e problem on differential grid. 
In Section \ref{subsec:LGa}, we show that the solutions of the Pad\'e problem give special solutions of the Lax equation and $q$-Garnier system.
 In Section \ref{subsec:sola}, we derive the explicit expressions of the special solutions 
in terms of the $q$-Appell Lauricella function \cite{Sakai05-2}.

Similarly, in Section \ref{sec:padeq} we study certain Pad\'e problem on $q$-grid and obtain
special solutions in terms of the generalized $q$-hypergeometric function. We note that 
the higher order $q$-Painlev\'e system given by Suzuki \cite{Suzuki15} also
has special solutions given in terms of the generalized $q$-hypergeometric function. 

%%%%%%%%%%%%%%%%

\section{A simple form of the $q$-Garnier system}\label{sec:Gar}

In this section, we give a reformulation of the evolution equation and the Lax equations of the $q$-Garnier system. In particular, the original Sakai's Lax form by $2 \times 2$ matrix is reformulated in a scalar form.

\subsection{Lax pair and the $q$-Garnier equation}\label{subsec:LG}

Fix a positive integer $N$ and a complex parameter $q$ ($0 < |q| <1$).
Let $a_1, \ldots, a_{N+1}$, $b_1, \ldots, b_{N+1}$, $c_1, c_2, d_1, d_2$ $\in \C^{\times}$ be
complex parameters  with a constraint $\prod_{i=1}^{N+1}\frac{a_i}{b_i}=q\prod_{i=1}^{2}\frac{c_i}{d_i}$ and $T_a: a \mapsto q a$ be the $q$-shift operator of parameter $a$. 

In this section, we put
\begin{equation}\label{eq:T}
T:=T_{a_1}^{-1}T_{b_1}^{-1},
\end{equation}
and the corresponding shifts are denoted as $\o{X}:=T(X)$ and $\u{X}:=T^{-1}(X)$. The operator $T$ plays the role of time evolution of the $q$-Garnier system. Though one can choose any $a_i, b_j$ instead of $a_1, b_1$, we consider the case $(i,j)=(1,1)$ for notational simplicity.

For an unknown function $y(x)$, we consider two linear difference equations: $L_2(x)=0$ between $y(x), y(qx), \o{y}(x)$ and $L_3(x)=0$ between $y(x), \o{y}(x), \o{y}(\frac{x}{q})$ defined as follows:
\begin{equation}\label{eq:L2L3} 
\begin{array}l
\ds L_2(x):=F(f,x)\o{y}(x)-A_1(x)y(qx)+(x-b_1)G(g,x)y(x),\\
\ds L_3(x):=F(\o{f},\frac{x}{q})y(x)+(x-a_1)G(g,\frac{x}{q})\o{y}(x)-qc_1c_2B_1(\frac{x}{q})\o{y}(\frac{x}{q}),
\end{array}
\end{equation}
where
\begin{equation}\label{eq:ABFG} 
\begin{array}l
\ds A(x)=\prod_{i=1}^{N+1}(x-a_i),\quad B(x)=\prod_{i=1}^{N+1}(x-b_i),\quad
\ds A_1(x)=\dfrac{A(x)}{x-a_1}, \quad B_1(x)=\dfrac{B(x)}{x-b_1},\\
\ds F(f,x)=\sum_{i=0}^N f_i x^i \quad G(g,x)=\sum_{i=0}^{N-1} g_i x^{i},
\end{array}
\end{equation}
 and $\o{f}_i=T(f_i)$, and $f_0, \ldots, f_N, g_0, \ldots, g_{N-1} \in\P^{1}$ are variables depending on parameters $a_i, b_i, c_i, d_i$. 

In this paper, we call $q^{\rho_0}$ (resp. $q^{\rho_{\infty}}$) the exponents at $x=0$ (resp. $x=\infty$) when  solutions $y(x)$ have the form 
\begin{equation}\label{eq:y0}
y(x)=k_0x^{\rho_0}(1+O(x)) \quad \mbox{at $x=0$},
\end{equation}
\begin{equation}\label{eq:yinf}
y(x)=k_{\infty}x^{\rho_{\infty}}(1+O\Big(\frac{1}{x}\Big)) \quad \mbox{at $x=\infty$}.
\end{equation}
\begin{prop}\label{prop:Geq}
We assume that  the exponents of solutions $y(x)$ of the linear equations $L_2(x)=0$ and $L_3(x)=0$ (\ref{eq:L2L3}) are $q^{\rho_0}=d_1,d_2$ and $q^{\rho_{\infty}}=c_1,c_2$. Then the compatibility of the equations $L_2(x)=0$ and $L_3(x)=0$ gives the following conditions:
%\begin{equation}\label{eq:y0}
%y(x)=k_0x^{\rho_0}(1+O(x)) \quad \mbox{at $x=0$},
%\end{equation}
%\begin{equation}\label{eq:yinf}
%y(x)=k_{\infty}x^{\rho_{\infty}}(1+O\Big(\frac{1}{x}\Big)) \quad \mbox{at $x=\infty$}
%\end{equation}
%where $q^{\rho_0}=d_1,d_2$ and $q^{\rho_{\infty}}=c_1,c_2$. Then the compatibility of $L_2$ and $L_3$ (\ref{eq:L2L3}) gives the following conditions:
\begin{equation}\label{eq:gd}
c_1c_2A_1(x)B_1(x)-(x-a_1)(x-b_1)G(g,x)G(\u{g},x)=0
\quad {\rm for} \quad F(f,x)=0,
\end{equation}
\begin{equation}\label{eq:fu1}
qc_1c_2A_1(x)B_1(x)-F(f,x)F(\o{f},x)=0
\quad {\rm for} \quad G(g,x)=0,
\end{equation}
\begin{equation}\label{eq:fu2}
f_N\o{f}_N=q (g_{N-1}-c_1)(g_{N-1}-c_2),\quad f_0\o{f}_0=a_1b_1(g_0-e_1)(g_0-e_2),
\end{equation}
where $e_i=\frac{d_i\nu}{a_1b_1}$ and $\nu=\prod_{i=1}^{N+1}(-a_i)$.
% and $d_1, d_2$ ($c_1, c_2$, resp.) is one of the exponents of the linear equation $L_1(x)$ at $x=0$ ($x=\infty$ resp.).
\end{prop}

\prf 
Under the condition $F(f,x)=0$, eliminating $y(x)$ and $y(qx)$ from $L_2(x)=\u{L}_3(qx)=0$, we obtain eq.(\ref{eq:gd}). Similarly, for $G(g,x)=0$, eliminating $y(qx)$ and $\o{y}(x)$ from $L_2(x)=L_3(qx)=0$, we have eq.(\ref{eq:fu1}).
%\begin{equation}\label{eq:fu3}
%qc_1c_2A_1(x)B_1(x)-F(f,x)F(\o{f},x)=0\quad {\rm for} \quad G(g,x)=0.
%\end{equation}
Considering exponents of the solutions $y(x)$ around $x=0$ and $x=\infty$, by the expansions (\ref{eq:y0}) and (\ref{eq:yinf}), we obtain eq.(\ref{eq:fu2}).
$\square$

%We remark that similar computations based on the contiguity type Lax pair have been done in \cite{Ikawa13, NTY13, Yamada14}. 

Though the conditions (\ref{eq:gd})$-$(\ref{eq:fu2}) are given as birational  equations for $2N+1$ variables $f_0$, $\ldots$, $f_N$, $g_0$,$\ldots$, $g_{N-1}$, they can be reduced to birational equations for $2N$ variables $\frac{f_1}{f_0}$,$\ldots$, $\frac{f_N}{f_0}, g_0$,$\ldots$, $g_{N-1}$ (see eqs.(\ref{eq:xu})$-$(\ref{eq:lu2}) in Section \ref{subsec:root}). In Section \ref{subsec:sufficiency}, eqs.(\ref{eq:gd})$-$(\ref{eq:fu2}) are proved to be sufficient for the compatibility of the equations $L_2(x)=0$ and $L_3(x)=0$ (\ref{eq:L2L3}).

The most fundamental object is the linear difference equation $L_1(x)=0$ between $y(qx)$, $y(x)$ and $y(\frac{x}{q})$. 
%We consider the following linear three term relation : $L_1(x)$ between $y(qx)$, $y(x)$, $y(x/q)$. 
%\begin{lem}
Eliminating $\o{y}(x)$ and $\o{y}(\frac{x}{q})$ from $L_2(x)= L_2(\frac{x}{q})=L_3(x)=0$ (\ref{eq:L2L3}), 
we have the linear equation $L_1(x)=0$, where
\begin{equation}\label{eq:L1}
\begin{array}{l}
\ds L_1(x):=A(x)F(f,\frac{x}{q})y(q x)+qc_1c_2B(\frac{x}{q})F(f,x)y(\frac{x}{q})\\[5mm]
\ds -\Big\{(x-a_1)(x-b_1)F(f,\frac{x}{q})G(g,x)+\dfrac{F(f,x)}{G(g,\frac{x}{q})}V(f,\o{f},\frac{x}{q})\Big\}y(x),
\end{array}
\end{equation}
and
\begin{equation}\label{eq:V}
V(f,\o{f},x)=qc_1c_2A_1(x)B_1(x)-F(f,x)F(\o{f},x).
\end{equation}
%\end{lem}
%As it is expected from (\ref{eq:L1}), 
%The following characterization of $L_1$ (\ref{eq:L1}) is important for the study of the compatibility.
\begin{lem}\label{lem:L1}
The equation $L_1(x)=0$ (\ref{eq:L1}) has the following properties: 

(i) It is a linear three term equation between $y(qx)$, $y(x)$ and $y(\frac{x}{q})$, and the coefficients of $y(qx)$, $y(x)$ and $y(\frac{x}{q})$ are polynomials of degree $2N+1$ in $x$,

(ii) The coefficient of $y(qx)$ (resp. $y(\frac{x}{q})$) has zeros at $x=a_1, \cdots , a_{N+1}$ (resp. $x=qb_1, \cdots , qb_{N+1}$),

(iii) The exponents of solutions $y(x)$ are $d_1, d_2$ (at $x=0$) and $c_1, c_2$ (at $x=\infty$),

(iv) The $N$ points $x$ such that $F(f,x)=0$ are apparent singularities (i.e., the solutions are regular there),  where 
\begin{equation}\label{eq:gdef1}
\dfrac{y(qx)}{y(x)}=\dfrac{G(g,x)(x-b_1)}{A_1 (x)} \quad {\mbox for}\quad F(f,x)=0
\end{equation}
holds. 

Conversely, the equation $L_1(x)=0$ (\ref{eq:L1}) is uniquely characterized by these properties (i)$-$(iv).
% once the coefficients of $y(q x)$ and $y(x/q)$ are given in the equation $L_1(x)$.
\end{lem}
\prf
The properties (i)$-$(iv) follows by computation using eqs.(\ref{eq:fu1}) and (\ref{eq:fu2}). The polynomiality of the coefficient of $y(x)$ follows from eq.(\ref{eq:fu1}). The converse can easily be confirmed by counting the number of free coefficients.  
$\square$

%\begin{dfn}\label{dfn:L1}
%The three term relation $L_1(x)$ is written in the form
%\begin{equation}\label{eq:L1}
%\begin{array}l
%\ds L_1(x):=A(x)F(f,\frac{x}{q})y(q x)-R_{2N+1}(x)y(x)+q c_1 c_2 B(\frac{x}{q})F(f,x)y(\frac{x}{q})=0,\\
%\ds A(x):=\prod_{j=1}^{N+1}(x-a_j),\quad \ds B(x):=\prod_{j=1}^{N+1}(x-b_j),\quad \ds F(f,x):=\sum_{j=0}^N f_j x^j,
%\end{array}
%\end{equation}
%where $f_0, \ldots, f_N$ are some variables depending on parameters $a_i,b_i,c_i,d_i$ but independent of $x$.
%The eq.(\ref{eq:L1}) satisfies the three conditions: (i) $R_{2N+1}(x)$ is a polynomial of degree $2N+1$, 
%(ii) the exponents are $d_1, d_2$ (at $x=0$) and $c_1, c_2$ (at $x=\infty$) with a constraint $\frac{1}{q}\prod_{j=1}^{N+1}\frac{a_j}{b_j}=\prod_{j=1}^{2}\frac{c_j}{d_j}$,
%(iii) the $N$ points $x$ : $F(f,x)=0$ are the apparent singularities (i.e., the solutions are regular there) such that 
%\begin{equation}\label{eq:gdef1}
%\dfrac{y(qx)}{y(x)}=\dfrac{G(g,x)(x-b_1)}{A_1 (x)} \quad {\mbox for}\quad F(f,x)=0.
%\end{equation}
%\end{dfn}

\subsection{Correspondence to Sakai's Lax form}\label{subsec:Sakai}

In \cite{Sakai05-1}, Sakai formulated the $q$-Garnier system as a multivariable extension of the sixth $q$-Painlev\'e equation, by using the connection preserving deformation of the following linear $q$-difference equation:
%For complex parameters $\alpha_i$, $\theta_j$, $\kappa_j$, $0 < |q| <1$ $\in \C^{\times}$ $(i=1,\ldots, N, j=1,2)$, 
\begin{equation}\label{eq:M1}
Y(qx)={\mathcal A}(x)Y(x), \quad {\mathcal A}(x):=\left[\begin{array}{cc} a(x)&b(x)\\ c(x)&d(x)\end{array}\right], \quad
Y(x):=\left[\begin{array}c \tilde{y}_1(x)\\ \tilde{y}_2(x) \end{array}\right].
\end{equation}

Let $\alpha_1 , \ldots \alpha_{2N+2} , \kappa_1 , \kappa_2 , \theta_1 , \theta_2 \in \C^{\times}$ be complex parameters. 
The coefficient matrix ${\mathcal A}(x)$ are defined by the following conditions:
(i) ${\mathcal A}(x)=\sum_{i=0}^{N+1}A_i x^i$, 
(ii) $A_{N+1}={\rm diag}(\kappa_1,\kappa_2)$ and $A_{0}$ has eigenvalues $\theta_1$ and $\theta_2$.
(iii) ${\rm det}{\mathcal A}(x)=\kappa_1\kappa_2\prod_{i=1}^{2N+2}(x-\alpha_i)$,
%where $\kappa_i, \theta_i, \alpha_i$ are given constants 
such that $\kappa_1\kappa_2\prod_{i=1}^{2 N+2}\alpha_i=\theta_1\theta_2$.
The conditions (i)$-$(iii) determine the matrix ${\mathcal A}(x)$ up to $2N+1$ free parameters. $2N$ of them are the dependent variables of the $q$-Garnier system, and one natural choice of them are given by variables $\{\lambda_i, \tilde{\mu}_i\}_{i=1}^N \in\P^{1}\times\P^{1}$ where $b(\lambda_i)=0$ and $\tilde{\mu}_i=a(\lambda_i)=\frac{\tilde{y}_1(q\lambda_i)}{\tilde{y}_1(\lambda_i)}$ (a kind of Sklyanin's "magic recipe", see \cite{Skl, SklTake} for example). We note that $\mu_i$ in \cite{Sakai05-1} is rewritten as $\tilde{\mu}_i$ here since $\mu_i$ is used for another variable (see Section \ref{subsec:root}). The remaining one free parameter (the normalization of the polynomial $b(x)$) is a gauge parameter.

The system (\ref{eq:M1}) can be equivalently described by the following scalar equation for the first component $\tilde{y}_1(x)$:
\begin{equation}\label{eq:S1}
b(\frac{x}{q})\tilde{y}_1(qx)-\{b(\frac{x}{q})a(x)+b(x)d(\frac{x}{q})\}\tilde{y}_1(x)+b(x){\rm det}{\mathcal A}(\frac{x}{q})\tilde{y}_1(\frac{x}{q})=0.
\end{equation}
Here $N$ points such that $b(x)=0$ ( i.e., $x=\lambda_i$) are apparent singularities. 

\begin{prop}\label{prop:LS1}
The linear equation $L_1(x)=0$ (\ref{eq:L1}) is equivalent to the linear equation (\ref{eq:S1}) up to a gauge transformation and changes of variables and parameters.
\end{prop}
\prf
By a gauge transformation: $\tilde{y}_1(x)=H(x)y_1(x)$ with $\frac{H(qx)}{H(x)}=\prod_{i=1}^{N+1}(x-\alpha_i)$, the system (\ref{eq:S1}) can be written as
\begin{equation}\label{eq:SS1}
\begin{array}{l}
\prod_{i=1}^{N+1}(x-\alpha_i)b(\frac{x}{q})y_1(qx)-\{b(\frac{x}{q})a(x)+b(x)d(\frac{x}{q})\}y_1(x)\\
+\kappa_1\kappa_2\prod_{i=N+2}^{2N+2}(\frac{x}{q}-\alpha_i)b(x)y_1(\frac{x}{q})=0.
\end{array}
\end{equation}
Then, eq.(\ref{eq:SS1}) has the following properties:

(i) It is a linear three term equation between $y_1(qx)$, $y_1(x)$ and $y_1(\frac{x}{q})$, and the coefficients of $y_1(qx)$, $y_1(x)$ and $y_1(\frac{x}{q})$ are polynomials of degree $2N+1$ in $x$, 

(ii) The coefficient of $y_1(qx)$ (resp. $y_1(\frac{x}{q})$) has zeros at $x=\alpha_1, \cdots , \alpha_{N+1}$ (resp. $x=q \alpha_{N+2}, \cdots , q \alpha_{2N+2}$). 

(iii) The exponents of the solutions $y_1(x)$ are $\frac{\theta_1}{\prod_{i=1}^{N+1}(-\alpha_i)}, \frac{\theta_2}{\prod_{i=1}^{N+1}(-\alpha_i)}$ (at $x=0$) and $\kappa_1, q^{-1}\kappa_2$ (at $x=\infty$), 

(iv) $N$ points $x=\lambda_i$ such that $b(x)=0$ are apparent singularities, where 
\begin{equation}\label{eq:mudef}
\frac{y_1(q\lambda_i)}{y_1(\lambda_i)}=\dfrac{\tilde{\mu}_i}{\prod_{i=1}^{N+1}(\lambda_i-\alpha_i)}
\end{equation}
holds. 

Conversely, eq.(\ref{eq:SS1}) is uniquely characterized by these properties (i)$-$(iv). Hence we see that the equation $L_1(x)=0$ (\ref{eq:L1}) is equivalent to eq.(\ref{eq:SS1}) up to changes of variables and parameters due to Lemma \ref{lem:L1}.
$\square$

\subsection{Sufficiency for the compatibility}\label{subsec:sufficiency}

Here we show that the birational equations (\ref{eq:gd})$-$(\ref{eq:fu2}) are sufficient for the compatibility of the equations $L_1(x)=0$ (\ref{eq:L1}) and $L_2(x)=0$ (or $L_3(x)=0$) (\ref{eq:L2L3}). Hence eqs.(\ref{eq:gd})$-$(\ref{eq:fu2}) can be regarded as the $q$-Garnier system.

To prove the sufficiency, we first study the linear difference equation $L_1^*(x)=0$ between $\o{y}(qx)$, $\o{y}(x)$ and $\o{y}(\frac{x}{q})$.  
%\begin{lem}
Eliminating ${y}(x)$ and  ${y}(qx)$ from $L_2(x)=L_3(x)=L_3(qx)=0$ (\ref{eq:L2L3}), we have the following expression:
\begin{equation}\label{eq:L1*}
\begin{array}{l}
\ds L_1^{*}(x):=\o{A}(x)F(\o{f},\frac{x}{q})\o{y}(q x)+qc_1c_2\o{B}(\frac{x}{q})F(\o{f},x)\o{y}(\frac{x}{q})\\[5mm]
\ds -\frac{1}{q}\Big\{(x-a_1)(x-b_1)F(\o{f},x)G(g,\frac{x}{q})+\dfrac{F(\o{f},\frac{x}{q})}{G(g,x)}V(f,\o{f},x)\Big\}\o{y}(x),
\end{array}
\end{equation}
where $V(f,\o{f},x)$ is given in eq.(\ref{eq:V}).
%\end{lem}

The following can be proved in the similar way as Lemma \ref{lem:L1}.

\begin{lem}\label{lem:L1*}
The equation $L_1^*(x)=0$ (\ref{eq:L1*}) has the following properties: 

(i) It is a linear three term equation between $\o{y}(qx)$, $\o{y}(x)$ and $\o{y}(\frac{x}{q})$,and the coefficients of $\o{y}(qx)$, $\o{y}(x)$ and $\o{y}(\frac{x}{q})$ are polynomials of degree $2N+1$ in $x$,

(ii) The coefficient of $\o{y}(qx)$ (resp. $\o{y}(\frac{x}{q})$) has zeros at $x=\frac{a_1}{q}, a_2 , \cdots , a_{N+1}$ (resp. $x=b_1, qb_2 , \cdots , qb_{N+1}$),

(iii) The exponents of solutions $\o{y}(x)$ are $d_1, d_2$ (at $x=0$) and $c_1, c_2$ (at $x=\infty$),

(iv) The $N$ points $x$ such that $F(\o{f},x)=0$ are apparent singularities, where 
\begin{equation}\label{eq:gdef2}
\dfrac{\o{y}(qx)}{\o{y}(x)}=\dfrac{B_1(x)}{c_1c_2G(g,x)(x-\frac{a_1}{q})} \quad {\mbox for}\quad F(\o{f},x)=0
\end{equation}
holds. 

Conversely, the equation $L_1^*(x)=0$ (\ref{eq:L1*}) is uniquely characterized by these properties (i)$-$(iv).
% once the coefficients of $y(q x)$ and $y(x/q)$ are given in the equation $L_1(x)$.
\end{lem}

\begin{thm}\label{thm:Lax}
The linear diference equations $L_1(x)=0$ (\ref{eq:L1}) and $L_2(x)=0$ (\ref{eq:L2L3}) are compatible
if and only if the birational equations (\ref{eq:gd})$-$(\ref{eq:fu2}) for variables $f_0, \ldots, f_N$ and $g_0, \ldots, g_{N-1}$ are satisfied. 
\end{thm}
\prf
The compatibility means that $T(L_1)=L_1^{\ast}$, i.e. the commutativity of  the following: 
\begin{equation}\label{eq:classification}\nonumber
\begin{array}{ccc}
L_1 \ (\mbox{Lemma} \  \ref{lem:L1})\quad &=&L_1\  (\ref{eq:L1})\\
&&\uparrow\\
\downarrow \mbox{$T$-shift}&&L_2 , L_3 \hspace{2mm}(\ref{eq:L2L3})\\
&&\downarrow\\
L_1^* \ (\mbox{Lemma} \  \ref{lem:L1*})\ &=& L_1^* \ (\ref{eq:L1*}).
\end{array}
\end{equation}
This can be checked by the characterizations of the equations $L_1=0$ and $L_1^*=0$ in Lemmas \ref{lem:L1} and \ref{lem:L1*}, and the relation: $T$(\ref{eq:gdef1})$=$ (\ref{eq:gdef2}) which follows from eq.(\ref{eq:gd}).
$\square$

\subsection{Expressions in terms of roots}\label{subsec:root}

%In this subsection, we will rewrite the expressions  (\ref{eq:L1}), (\ref{eq:L2L3}), (\ref{eq:gd})$-$(\ref{eq:fu2}) of section \ref{subsec:LG} into more explicit forms. We will omit the proofs because it is only the change of variables from the coefficients $\{f_i, g_i\}$ to the roots $\{\lambda_i, \xi_i\}$ of polynomials $F(f,x)$ and $G(g,x)$.

Introduce variables $\lambda_i$ and $\mu_i$ ($i=1, \ldots, N$) $\in\P^{1}$ such that $F(f,x)=f_N\prod_{i=1}^{N}(x-\lambda_i)=:\Lambda(x)$
and $\mu_i:=\frac{y(q\lambda_i)}{y(\lambda_i)}$.
By the definition of $\mu_i$ and eq.(\ref{eq:gdef1}), the variables $\{\mu_i\}$ are related to $\{g_i\}$ as $\mu_i=\frac{(\lambda_i-b_1)G(g,\lambda_i)}{A_1(\lambda_i)}$. 
Using the characterization in Lemma \ref{lem:L1} and the partial fraction expansion, one can rewrite linear three term equation $L_1(x)=0$ (\ref{eq:L1}) as
\begin{equation}\label{eq:L1root}
\begin{array}{l}
\ds \dfrac{L_1(x)}{x\Lambda(x)\Lambda(\frac{x}{q})}=\dfrac{A(x)}{x\Lambda(x)}y(q x)
+\dfrac{qc_1c_2B(\frac{x}{q})}{x\Lambda(x/q)}y(\frac{x}{q})\\[5mm]
\ds -\Big[\frac{\nu(d_1+d_2)}{f_0x}+\dfrac{c_1+c_2}{f_N}+\displaystyle\sum_{i=1}^{N}\frac{1}{\lambda_i\Lambda^{\prime}(\lambda_i)}\Big(
\frac{A(\lambda_i)\mu_i}{x-\lambda_i}+\frac{qc_1c_2B(\lambda_i)}{(x-q \lambda_i)\mu_i}\Big)\Big]y(x)=0,
\end{array}
\end{equation}
where $ f_0=f_N\prod_{i=1}^N(-\lambda_i)$ and $\nu=\prod_{i=1}^{N+1}(-a_i)$.

\begin{rem}{\bf On the characterization in terms of  variables $(\lambda_i, \mu_i)$} The characterization in Lemma \ref{lem:L1} is that for the polynomial in $x$, however we have a similar characterization as a curve in the coordinates $(\lambda_i, \mu_i)$. Namely, in case of $i=1$ for example, the equation $x\Lambda(x)\Lambda(\frac{x}{q})L_1=0$ has the following characterization in $(\lambda_1, \mu_1)$: (i) It is a polynomial equation of bidegree $(N+2,2)$, 
(ii) passing through the following $3N+9$ points $(0,d_i)_{i=1}^2$, $(\infty,c_i)_{i=1}^2$, 
$(b_i,0)_{i=1}^{N+1}$, $(a_i,\infty)_{i=1}^{N+1}$, $(x,0)$, $(\frac{x}{q},\infty)$, $(x,\frac{y(q x)}{y(x)})$,
$(\frac{x}{q}, \frac{y(x)}{y(\frac{x}{q})})$, $(\lambda_i,\mu_i)_{i=2}^N$ $(i \ne1)$. 
%By the symmetry, there exist same characterizations for the other variables $(\lambda_i, \mu_i)$ also. 
This characterization is a generalization of those of Lax equations $L_1$ for the discrete Painlev\'e equations as a curve of bidegree $(3,2)$ passing through $12$ points \cite{KNY15, Yamada09-2, Yamada11}.
\end{rem}

The linear three term equations $L_2(x)=0$ and $L_3(x)=0$ (\ref{eq:L2L3}) can also be rewritten as
\begin{equation}\label{eq:L2L3root}
\begin{array}l
\ds L_2(x)=\Lambda(x)\o{y}(x)-A_1(x)y(qx)+(x-b_1)\Xi(x)y(x)=0,\\[3mm]
\ds L_3(x)=\o{\Lambda}(\frac{x}{q})y(x)+(x-a_1)\Xi(\frac{x}{q})\o{y}(x)-qc_1c_2B_1(\frac{x}{q})\o{y}(\frac{x}{q})=0,
\end{array}
\end{equation}
%where, $A_1(x), B_1(x), r$ are the same as proposition \ref{prop:L2L3}, and 
where $\Xi(x)=G(g,x)=g_{N-1}\prod_{i=1}^{N-1}(x-\xi_i)$ with the roots $\xi_i$ of the polynomial $\Xi(x)$. Then the evolution equation (\ref{eq:gd})$-$(\ref{eq:fu2}) can be rewritten as
\begin{equation}\label{eq:xu}
\Xi(\lambda_i) \u{\Xi}(\lambda_i)
=c_1c_2 \dfrac{A_1(\lambda_i)B_1(\lambda_i)}{(\lambda_i-a_1)(\lambda_i-b_1)} \quad (i=1,\ldots,N),  
\end{equation}
\begin{equation}\label{eq:lu1}
\dfrac{\Lambda(\xi_i)\o{\Lambda}(\xi_i)}{f_N\o{f}_N}=c_1c_2\dfrac{A_1(\xi_i)B_1(\xi_i)}{(g_{N-1}-c_1)(g_{N-1}-c_2)} \quad (i=1,\ldots,N-1),
\end{equation}
\begin{equation}\label{eq:lu2}
\displaystyle \prod_{i=1}^{N} \lambda_i \o{\lambda}_i=\dfrac{a_1b_1}{q}\dfrac{(g_0-e_1)(g_0-e_2)}{(g_{N-1}-c_1)(g_{N-1}-c_2)},
\end{equation}
where $g_{0}=g_{N-1}\prod_{i=1}^{N-1}(-\xi_i)$. The evolution equation (\ref{eq:xu})$-$(\ref{eq:lu2}) is the $q$-Garnier equation in terms of $2N$ variables $\lambda_1,\ldots,\lambda_N,\xi_1,\ldots,\xi_{N-1}, g_0$ (or $g_{N-1}$). It is more explicit than eqs.(\ref{eq:gd})$-$(\ref{eq:fu2}) but nonbirational.
%and $e_i$ is the same as eq.(\ref{eq:fu2}).

\subsection{Reduction to the $q$-Painlev\'e equations}\label{subsec:dege}

We give a few comments on lower cases $N=1,2,3$.
In \cite{Sakai05-1}, the $q$-Painlev\'e equation of type $D_5^{(1)}$ has appeared as a case for the $q$-Garnier system with $N=1$. 
This is easily seen from eqs.(\ref{eq:gd})$-$(\ref{eq:fu2}).
For $N=2$ case, it is known \cite{Sakai06} that the $q$-Painlev\'e equation of type $E_6^{(1)}$ appears as a particular case for the $q$-Garnier system with $N=2$. In fact, we have

\begin{prop}\label{prop:E6}(\cite{Sakai06}) For the case $N=2$ with a constraint $c_1=c_2$, the $q$-Garnier equation (\ref{eq:gd})$-$(\ref{eq:fu2}) admits the
following reduction 
\begin{equation}\label{eq:E6}
\begin{array}l
(fg-1)(f\u{g}-1)=\dfrac{(1-a_2f)(1-a_3f)(1-b_2f)(1-b_3f)}{(1-a_1f)(1-b_1f)},\\
%\end{array}
%\end{equation}
%\begin{equation}\label{eq:E6fu}
%\begin{array}l
(fg-1)(f\o{g}-1)=\dfrac{(1-\frac{g}{a_2})(1-\frac{g}{a_3})(1-\frac{g}{b_2})(1-\frac{g}{b_3})}{(1-\frac{c_1g}{e_1})(1-\frac{c_1g}{e_2})},
\end{array}
\end{equation}
where $e_i=\frac{d_i\nu}{a_1b_1}$ and $\nu=\prod_{i=1}^{N+1}(-a_i)$. 
\end{prop}
\prf
Under the case $N=2$ with the constraint, eqs.(\ref{eq:gd})$-$(\ref{eq:fu2}) admit a specialization $f_2=0$ and $g_1=c_1$. Then we obtain the results where $f=-\frac{f_1}{f_0}$ and $g=-\frac{g_0}{c_1}$.
$\square$ 

Eq.(\ref{eq:E6}) is the $q$-Painlev\'e equation of type $E_6^{(1)}$ \cite{KNY15, Sakai01}.

For $N=3$ case, the $q$-Painlev\'e equation of type $E_7^{(1)}$ appears as a particular case for the $q$-Garnier system with $N=3$. In fact, we have
\begin{prop}\label{prop:E7}
For the case $N=3$ with constraints $d_1=d_2$ and $c_1=c_2$,  the $q$-Garnier equation (\ref{eq:gd})$-$(\ref{eq:fu2}) admits the
following reduction 
\begin{equation}\label{eq:E7}
\begin{array}l
\left\{g+(f+\frac{e_1}{c_1f})\right\}
\left\{\u{g}+(f+\frac{qe_1}{c_1f})\right\}=\dfrac{(f-a_2)(f-a_3)(f-a_4)(f-b_2)(f-b_3)(f-b_4)}{f^2(f-a_1)(f-b_1)},\\
%\end{array}
%\end{equation}
%\begin{equation}\label{eq:E7fu}
%\begin{array}l
\dfrac{(1-x_1/f)(1-x_1/\o{f})}{(1-x_2/f)(1-x_2/\o{f})}=\dfrac{x_2^2(x_1-a_2)(x_1-a_3)(x_1-a_4)(x_1-b_2)(x_1-b_3)(x_1-b_4)}{x_1^2(x_2-a_2)(x_2-a_3)(x_2-a_4)(x_2-b_2)(x_2-b_3)(x_2-b_4)},
\end{array}
\end{equation}
where $e_i=\frac{d_i\nu}{a_1b_1}$, $\nu=\prod_{i=1}^{N+1}(-a_i)$ and $x=x_1, x_2$ are solutions of the equation $g+(x+\frac{e_1}{c_1x})=0$. 
\end{prop}
\prf
Under the case $N=3$ with the constraints, eqs.(\ref{eq:gd})$-$(\ref{eq:fu2}) admit a specialization $f_0=f_3=0$, $g_0=e_1$ and $g_2=c_1$. Then we obtain the results where $f=-\frac{f_1}{f_2}$ and $g=\frac{g_1}{c_1}$.
$\square$ 

Eq.(\ref{eq:E7}) for the variables $f$ and $g$ is a kind of the $q$-Painlev\'e equation of type $E_7^{(1)}$, but the direction of the time evolution is different from the standard one given in \cite{KNY15,Sakai01}. The relation of them will be discussed in \cite{NYyokoku}. 

\subsection{Correspondence of parameters and variables in \S2, \S4, \S5}\label{subsec:234} 

In this paper, parameters $a_1, \ldots, a_{N+1}$, $b_1, \ldots, b_{N+1}$, $c_1, c_2$, $d_1, d_2$, $m, n$ and variables $f_0, \ldots, f_{N}$, $g_0,\ldots,g_{N-1}$ are used in slightly different means in \S\ref{sec:Gar}, \S\ref{sec:padea} and \S\ref{sec:padeq}. Their relations are given as follows:
\begin{equation}\label{eq:notation}
\begin{array}l
(a_i, b_i)^{\S2}=(\frac{1}{a_i},\frac{1}{b_i})^{\S4}=(\frac{1}{a_i},\frac{1}{b_i})^{\S5} \quad {\mbox (i=1,\ldots,N)}, \\
(a_{N+1},b_{N+1})^{\S2}=(\frac{1}{a_{N+1}},\frac{1}{b_{N+1}})^{\S4}=(q^{m+n},\frac{1}{q})^{\S5},\\%[3mm]
(c_1,c_2)^{\S2}=(q^m, q^n\prod_{j=1}^{N+1}\frac{b_j}{a_j})^{\S4}=(q^m, cq^n\prod_{j=1}^{N}\frac{b_j}{a_j})^{\S5}, \\%[3mm]
(d_1,d_2)^{\S2}=(1, q^{m+n+1})^{\S4}=(1,c)^{\S5},\\%[3mm]
f_0^{\S2}=(\frac{-a_1}{\nu}(1-g_0))^{\S4}=(\frac{-a_1}{\nu}(1-g_0))^{\S5},\\
f_i^{\S2}=(\frac{-a_1}{\nu}(1-g_0)f_i)^{\S4}=(\frac{-a_1}{\nu}(1-g_0)f_i)^{\S5} \quad {\mbox (i=1,\ldots,N)},\\
\o{f}_0^{\S2}=(\frac{-b_1q^{m+n+1}}{\nu}(1-\frac{g_0}{q^{m+n+1}}))^{\S4}=(\frac{-b_1c}{\nu}(1-\frac{g_0}{c}))^{\S5},\\%[3mm]
\o{f}_i^{\S2}=(\frac{-b_1q^{m+n+1}}{\nu}(1-\frac{g_0}{q^{m+n+1}})\o{f}_i)^{\S4}=(\frac{-b_1c}{\nu}(1-\frac{g_0}{c})\o{f}_i)^{\S5} \quad {\mbox (i=1,\ldots,N)},\\%[3mm]
%w_0^{\S4}=1-(\frac{g_0}{d_1})^{\S4} \quad w_1^{\S4}=1-(\frac{g_0}{d_2})^{\S4},\quad w_0^{\S5}=1-(\frac{g_0}{d_1})^{\S5} \quad w_1^{\S5}=1-(\frac{g_0}{d_2})^{\S5},\\
g_i^{\S2}=(\frac{a_1b_1}{\nu}g_i)^{\S4}=(\frac{a_1b_1}{\nu}g_i)^{\S5}\quad (i=0,\ldots,N-1)
\end{array}
\end{equation}
where $\nu=\prod_{i=1}^{N+1}(-a_i)$.

%section
\section{Autonomous case}\label{sec:QRT}
In this section, we define a generalization of the QRT system \cite{QRT89} (see also \cite{Tsuda04}) for hyperelliptic curves
and discuss its relation to the $q$-Garnier system.

\subsection{Generalization of the QRT map for hyperelliptic curve}\label{subsec:QRT}

\noindent
Let $C$ be a curve of bidegree $(N+1,2)$ on $\P^1\times\P^1$
which passes through given $2N+5$ points $P_1, \ldots, P_{2N+5}$.
The number of free parameters of the defining polynomial is $3(N+2)-(2N+5)=N+1$, hence the curve $C$ forms an $N$ dimensional family, generically of genus $N$.

The dynamical variables of the generalized QRT mapping is a set (a divisor) of $N$ points $\{Q_1, \ldots, Q_N\}$ on the curve $C$. Following Mumford \cite{Mum84} we represent it by a pair of functions $\Phi(x)$ and $\Psi(x)$, where $\Phi(x)$ is a polynomial of degree $N$ and $\Psi(x):=\frac{S(x)}{R(x)}$ is a rational function of degree $N$, such that $Q_i=(x_i,\Psi(x_i))$, $\Phi(x_i)=0$, $(i=1,\ldots, N)$. Note that the normalization of $\Phi(x)$ is irrelevant. 
A generalized QRT map is defined as follows:

(1) Fix $N$ free parameters of the curve $C$ so that it passes the initial points $Q_1, \ldots, Q_N$.
%; namely $\varphi(x, \Psi(x))$ is divisible by $\Phi(x)$. 
We represent the resulting curve as $C_0: \varphi(x,y):=\alpha(x)y^2+\beta(x)y+\gamma(x)=0$.

(2) Take a subset of indices $I \subset \{1, \ldots, 2N+5\}$ with $|I|=N+1$ and
determine the rational function $\Psi(x)=\frac{S(x)}{R(x)}$ uniquely by the condition that the curve $y=\Psi(x)$ passes through the points $P_{i}=(x_{P_i}, y_{P_i}) (i \in I)$ and $Q_1, \cdots, Q_N$.
By definition 
$R(x)^2\varphi (x,\Psi(x))$ is divisible by $\prod_{i \in I}(x-x_{P_i})\Phi(x)$ and we can define an involution $\iota_x^I: (\Phi,\Psi)\mapsto (\tilde{\Phi},\Psi)$ 
by the relation
\begin{equation}\label{eq:x-flip}
R(x)^2 \varphi(x,\Psi(x))=\prod_{i \in I}(x-x_{P_i}) \Phi(x) \tilde{\Phi}(x).
\end{equation}

(3) Since the polynomial $\varphi(x,y)$ is of degree 2 in $y$, the other involution $\iota_y: (x,y)\mapsto (x,\tilde{y})$ can be defined simply as $y\tilde{y}=\frac{\gamma(x)}{\alpha(x) }$. Namely we have $\iota_y: (\Phi,\Psi) \mapsto (\Phi,\tilde{\Psi})$ where
\begin{equation}\label{eq:y-flip}
\Psi(x) \tilde{\Psi}(x)\alpha(x)= \gamma(x), \quad {\rm for} \ \Phi(x)=0.
\end{equation}

(4) We have the generalized QRT map defined by the iteration $T_I:=\iota_y \iota_x^{I}$ or $T_I^{-1}:=\iota_x^{I}\iota_y$. 
They are commutative: $T_I T_J=T_J T_I$ since they are translations on the Jacobian of the curve $C_0$.

\subsection{Relation to the $q$-Garnier system}

In order to apply the algorithm in Section \ref{subsec:QRT} to the $q$-Garnier system (\ref{eq:gd})$-$(\ref{eq:fu2}), we consider the following case. The points $P_1, \ldots, P_{2N+5}$ are taken as intersection points of the curve $C_0$ and four lines \footnote{A similar construction is possible for other curve of bidegree (2,2)
instead of the four lines.}: $x=0$, $x=\infty$, $y=0$ and $y=\infty$. Then 
%for complex parameters $a_1,\dots,a_{N+1}$, $b_1,\ldots,b_{N+1}$, $c_1,c_2$, $d_1,d_2$ $\in \C^{\times}$, 
we have one additional point $P_{2N+6}$ and the $2N+6$ points can be parametrized as $(a_i,\infty)_{i=1}^{N+1}$, $(b_i,0)_{i=1}^{N+1}$, $(\infty,c_i)_{i=1}^2$ and $ (0,d_i)_{i=1}^2$ with a constraint
$\prod_{i=1}^{N+1}\frac{a_i}{b_i}=\prod_{i=1}^2\frac{c_i}{d_i}$. 
Therefore the curve $C_0: \varphi(x,y)=0$ of bidegree $(N+1,2)$ can be written as
\begin{equation}
\begin{array}c
\ds \varphi(x,y):=A(x)y^2 -U(x)y+c_1c_2B(x)=0, \\
\ds \quad U(x):=\nu(d_1+d_2)+\sum_{i=1}^N u_i x^i+(c_1+c_2) x^{N+1}.
\end{array}
\end{equation}
Here $A(x)=\prod_{i=1}^{N+1}(x-a_i), B(x)=\prod_{i=1}^{N+1}(x-b_i)$ and $\nu=\prod_{i=1}^{N+1}(-{a_i})$ as in Section \ref{subsec:LG}. 
Note that the lowest/highest terms of $\varphi(x,y)$ in $x$ are given by $\varphi|_{x^0}=\nu(y-d_1)(y-d_2)$ and
$\varphi|_{x^{N+1}}=(y-c_1)(y-c_2)$.
The parameters (conserved quantities) $u_1, \cdots, u_N$ are determined by the condition $\varphi(Q_i)=0$ for the initial points: $Q_1, \cdots, Q_N$.

To adjust the formulation given above to that in Section \ref{sec:Gar}, we take the index set $I$ as $\{P_i | i \in I\}=\{(a_i,\infty)_{i=2}^{N+1}, (b_1,0)\}$, and put $\Phi(x):=F(x)$ and $\Psi(x):=\frac{(x-b_1)G(x)}{A_1(x)}$  where $F(x)=\sum_{i=0}^N f_i x^i$ and $G(x)=\sum_{i=0}^{N-1}g_i x^i$ as in eq.(\ref{eq:ABFG}). Then, the $\iota^I_x$-flip  defined by eq.(\ref{eq:x-flip}) takes the form
\begin{equation}
(x-a_1)(x-b_1)G(x)^2-U(x)G(x)+c_1c_2A_1(x)B_1(x)=F(x)\tilde{F}(x),
\end{equation}
where $A_1(x)=\frac{A(x)}{x-a_1}$ and $B_1(x)=\frac{B(x)}{x-b_1}$ as in eq.(\ref{eq:ABFG}). 
This relation determines the polynomial $\tilde{F}(x)$  by 
\begin{equation}\label{eq:QRT-f1}
F(x)\tilde{F}(x)= c_1c_2A_1(x)B_1(x), \quad {\rm for} \ G(x)=0, 
\end{equation}
%and additional conditions for the lowest/highest coefficients:
\begin{equation}\label{eq:QRT-f2}
f_N{\tilde{f_N}}=(g_{N-1}-c_1)(g_{N-1}-c_2),\quad f_0 \tilde{f_0}=a_1b_1(g_0-e_1)(g_0-e_2),
\end{equation}
where $e_i=\frac{d_i\nu}{a_1b_1}$ as in Proposition \ref{prop:Geq}. On the other hand, the $\iota_y$-flip (\ref{eq:y-flip}) gives
\begin{equation}\label{eq:QRT-g1}
(x-a_1)(x-b_1) G(x)\tilde{G}(x)= c_1c_2A_1(x)B_1(x), \quad {\rm for} \ F(x)=0.
\end{equation}

It is easy to see that
\begin{prop}
The birational equations (\ref{eq:QRT-f1})$-$(\ref{eq:QRT-g1}) correspond to the autonomous ($q=1$) version of the $q$-Garnier system (\ref{eq:gd})$-$(\ref{eq:fu2}).
\end{prop}

%\noindent
%{\bf Remark.} For a rational function $h=h(x)$,  define a transformation  $R(h) : (F(x),G(x)) \mapsto (F(x), \tilde{G}(x))$ by
%\begin{equation}
%\tilde{G}(x)\equiv h(x)G(x), \quad {\rm mod} \ F(x).
%\end{equation}
%Then the maps $T_{i,j}$ and $T_{k,l}$ are commutative in the sense that
%\begin{equation}
%R(h)\circ T_{i,j}\circ R(h^{-1})\circ T_{k,l}=T_{k,l}\circ R(h)\circ T_{i,j}\circ R(h^{-1}), \quad h=\dfrac{(x/a_i,x/b_j)_1}{(x/a_k,x/b_l)_1}.
%\end{equation}

\section{The Pad\'e problem on differential grid}\label{sec:padea}
In this section, we study certain Pad\'e approximation problem and solve it explicitly. As a result, we obtain some special solutions of the $q$-Garnier system given in terms of the $q$-Appell Lauricella function. 

\subsection{Lax pair and the $q$-Garnier equation}\label{subsec:LGa}

In this subsection, starting the Pad\'e approximation problem (\ref{eq:pade1a}), we derive linear difference relations (\ref{eq:L2L3a}) 
%and (\ref{eq:L1a}) 
and nonlinear relations (\ref{eq:gda})$-$(\ref{eq:fu2a}). 

For complex parameters $a_i \ldots, a_{N+1}$, $ b_1, \ldots, b_{N+1}$ $\in \C^{\times}$, we put
\begin{equation}\label{eq:psia}
\psi(x):=\prod_{i=1}^{N+1}\dfrac{(a_i x)_\infty}{(b_i x)_\infty}.
\end{equation}
Here and in what follows, we use the standard $q$-Pochhammer symbols defined as
\begin{equation}
(z)_\infty:=\prod_{i=0}^\infty (1-q^i z), \quad
(z)_s:=\frac{(z)_\infty}{(zq^s)_\infty},  \quad
(z_1,z_2, \ldots, z_k)_s:=(z_1)_s(z_2)_s \ldots (z_k)_s.
\end{equation}

Define polynomials $P(x)$ and $Q(x)$ of degree $m$ and $n$ $\in \Z_{\ge 0}$ by the following Pad\'e approximation condition:
\begin{equation}\label{eq:pade1a}
\psi(x)=\dfrac{P(x)}{Q(x)}+O(x^{m+n+1}).
\end{equation}
%This is the Pad\'e problem which is the bases of our study. 
Here the common normalizations of the polynomials $P(x)$ and $Q(x)$ are fixed as $P(0)=1$ tentatively. 

%we introduce a discrete transformation $T$ of the parameters and study the contiguity relations between the original and shifted objects. We remark that similar computations in this subsection have been already done in \cite{Ikawa13, Nagao15-2}. 

By the change of parameters (\ref{eq:notation}), the shift $T$ (\ref{eq:T}) becomes the following form
\begin{equation}\label{eq:Ta}
T=T_{a_1}T_{b_1}.
\end{equation}
%For any object $X$, the corresponding shifts are denoted as $\o{X}:=T(X)$, $\u{X}:=T^{-1}(X)$. 
%For example, the symbols $\o{P}(x)$, $\o{Q}(x)$ represent the polynomials defined by 
%\begin{equation}\label{eq:pade2a}
%\o{\psi}(x)=\dfrac{\o{P}(x)}{\o{Q}(x)}+O(x^{m+n+1}).
%\end{equation}
%We call the operators $T$ as the "time evolution", since it will play the role of time evolution of the $q$-Garnier system.

Let us consider two linear difference relations: $L_2(x)=0$ between $y(x), y(qx), \o{y}(x)$ and $L_3(x)=0$ between $y(x), \o{y}(x), \o{y}(\frac{x}{q})$ satisfied by the functions $y(x)=P(x)$ and $y(x)=\psi(x)Q(x)$. The following proposition shows that these relations are regarded as the Lax equations $L_2(x)=0$ and $L_3(x)=0$  for the $q$-Garnier system (cf. eq.(\ref{eq:L2L3})).

\begin{prop}\label{prop:L2L3a}
The linear relations $L_2(x)=0$ and $L_3(x)=0$ can be written as follows:
\begin{equation}\label{eq:L2L3a} 
\begin{array}l
\ds L_2(x)=(g_0)_1F(f,x)\o{y}(x)-A_1(x)y(qx)+(b_1 x)_1G(g,x)y(x)=0,\\
\ds L_3(x)=(\frac{g_0}{q^{m+n+1}})_1 F(\o{f},\frac{x}{q})y(x)+\frac{1}{q^{m+n+1}}(a_1 x)_1G(g,\frac{x}{q})\o{y}(x)-B_1(\frac{x}{q})\o{y}(\frac{x}{q})=0,
\end{array}
\end{equation}
where
\begin{equation}
\begin{array}l
\ds A(x)=\prod_{i=1}^{N+1}(a_i x)_1,\quad \ds B(x)=\prod_{i=1}^{N+1}(b_i x)_1,\quad \ds F(f,x)=1+\sum_{i=1}^N f_i x^i,\\
A_1(x)=\dfrac{A(x)}{(a_1 x)_1}, \quad B_1(x)=\dfrac{B(x)}{(b_1 x)_1}, \quad \ds G(g,x)=\sum_{i=0}^{N-1} g_i x^{i}.
\end{array}
\end{equation}
Here $\o{f}_i=T(f_i)$, and $f_1, \ldots, f_N, g_0, \ldots, g_{N-1}\in\P^{1}$ are some constants depending on parameters $a_i,b_i,m,n$. 
\end{prop}

\prf
%The proof is similar to the proof of Theorem 2.1 of \cite{Ikawa13} (se also \cite{Nagao15-2}). 
By the definition of the linear relations $L_2(x)=0$ and $L_3(x)=0$, they can be written as
{\small
\begin{align}\label{eq:L2L3matrixa}
L_2(x)\propto
\begin{vmatrix}
y(x) & y(qx) & \o{y}(x) \\
P(x) & P(qx) & \o{P}(x)\\
\psi(x)Q(x) & \psi(qx)Q(qx) & \o{\psi}(x)\o{Q}(x)
\end{vmatrix}=0, \\
L_3(x)\propto
\begin{vmatrix} 
y(x) & \o{y}(x) & \o{y}(\frac{x}{q}) \\
P(x) & \o{P}(x) & \o{P}(\frac{x}{q})\\
\psi(x)Q(x) & \o{\psi}(x)\o{Q}(x) & \o{\psi}(\frac{x}{q})\o{Q}(\frac{x}{q})
\end{vmatrix}=0.
\end{align}
}
Setting ${\bf y}(x):=\left[\begin{array}{c}P(x)\\ \psi(x)Q(x)\end{array}\right]$, define Casorati determinants $D_1(x)$, $D_2 (x)$ and $D_3(x)$ by
\begin{equation}\label{eq:Ddefa}
\begin{array}{l}
D_1(x):=\det[{\bf y}(x),{\bf y}(qx)],\quad 
D_2(x):=\det[{\bf y}(x),{\o{\bf y}}(x)],\quad D_3(x):=\det[{\bf y}(qx),\o{{\bf y}}(x)].
\end{array}
\end{equation}
Then, the linear relations $L_2(x)=0$ and $L_3(x)=0$ take the following forms:
\begin{equation}\label{eq:L2L3Da}
\begin{array}{l}
L_2(x)\propto D_1(x) \o{y}(x)-D_2(x)y(qx)+D_3(x)y(x)=0,
\\
L_3(x)\propto \o{D}_1(\frac{x}{q})y(x)+D_3(\frac{x}{q}) \o{y}(x)-D_2(x)\o{y}(\frac{x}{q})=0.
\end{array}
\end{equation}
The determinants (\ref{eq:Ddefa}) can be computed by the condition (\ref{eq:pade1a}) and the relations
\begin{equation}
\dfrac{\psi(qx)}{\psi(x)}=\prod_{i=1}^{N+1}\dfrac{(b_i x)_1}{(a_i x)_1},
\quad \dfrac{\o{\psi}(x)}{\psi(x)}=\dfrac{(b_1 x)_1}{(a_1 x)_1}.
\end{equation}
The results are given as 
\begin{equation}\label{eq:Da}
\begin{array}{l}
D_1(x)=\dfrac{\psi(x)}{A(x)}\left\{B(x)P(x)Q(qx)-A(x)P(qx)Q(x)\right\}=:w_0\dfrac{\psi(x)x^{m+n+1}}{A(x)}F(f,x),\\
D_2(x)=\dfrac{\psi(x)}{(a_1 x)_1}\left\{(b_1 x)_1P(x)\o{Q}(x)-(a_1 x)_1\o{P}(x)Q(x)\right\}=:w_1\dfrac{\psi(x)x^{m+n+1}}{(a_1 x)_1},\\
D_3(x)=\dfrac{\psi(x)}{A(x)}\left\{(b_1 x)_1A_1(x)P(qx)\o{Q}(x)-B(x)\o{P}(x)Q(qx)\right\}=:w_1\dfrac{\psi(x)x^{m+n+1}}{A(x)}(b_1 x)_1G(g,x),
\end{array}
\end{equation}
with some constants $w_0$ and $w_1$ depending on parameters $a_i$, $b_i$, $m$ and $n$.
% but independent of $x$. %Substituting eqs.(\ref{eq:Da}) into eq.(\ref{eq:L2L3Da}),
%and using the suitable transformation of $y(x)$ in $L_2$ i.e. $y(x)\mapsto Gy(x)$, $G/\o{G}=1/c$, $w=r/c^2$, 
%we obtain eq.(\ref{eq:L2L3a}), where 
The constants $w_0$ and $w_1$ are fixed as $w_0=(g_0)_1$ and $w_1=(\frac{g_0}{q^{m+n+1}})_1$ by the condition that eq.(\ref{eq:L2L3Da}) has a solution such as $y(0)=P(0)=1$.
$\square$

%We will derive the birational evolution equations among $f_i, g_i, \o{f}_i, \u{g}_i$ from the compatibility of $L_2, L_3$ (\ref{eq:L2L3}).
The following proposition can be proved in the similar way as Proposition \ref{prop:Geq}.
\begin{prop}\label{prop:Geqa}
The constants $f_1, \ldots ,f_N$ and $g_0, \ldots ,g_{N-1}$ satisfy the following relations:
%The compatibility of the relations $L_2$ and $L_3$ (\ref{eq:L2L3a}) gives the following condition:
\begin{equation}\label{eq:gda}
A_1(x)B_1(x)-\frac{1}{q^{m+n+1}}(a_1 x, b_1 x)_1G(g,x)G(\u{g},x)=0
\quad {\rm for} \quad F(f,x)=0,
\end{equation}
\begin{equation}\label{eq:fu1a}
A_1(x)B_1(x)-(g_0, \frac{g_0}{q^{m+n+1}})_1F(f,x)F(\o{f},x)=0
\quad {\rm for} \quad G(g,x)=0,
\end{equation}
\begin{equation}\label{eq:fu2a}
(g_0, \frac{g_0}{q^{m+n+1}})_1f_N\o{f}_N=\Big(\frac{a_1g_{N-1}}{q^{m+n}}+\frac{\prod_{i=2}^{N+1}(-b_i)}{q^m}\Big)\Big(b_1 g_{N-1}+q^m\prod_{i=2}^{N+1}(-a_i)\Big),
\end{equation}
%where $q^{m+n+1}$ (resp. $q^m$) is one of the exponents of the linear equation $L_1(x)$ (\ref{eq:L1a}) at $x=0$ (resp. $x=\infty$). 
\end{prop}
These relations (\ref{eq:gda})$-$(\ref{eq:fu2a}) are regarded as the $q$-Garnier system (cf. eqs.(\ref{eq:gd})$-$(\ref{eq:fu2})).

\subsection{Special solutions}\label{subsec:sola}

We derive the explicit forms (\ref{eq:F1a})$-$(\ref{eq:G2a}) of variables $\{f_i, g_i\}$ appearing in the Casorati determinants $D_1(x)$ and $D_3(x)$ (\ref{eq:Da}). They are interpreted as the special solutions for the $q$-Garnier system (\ref{eq:gda})$-$(\ref{eq:fu2a}) due to Proposition \ref{prop:Geqa}.

\begin{prop}\label{prop:PQ1a}
For any given function $\psi(x)=\sum_{k=0}^{\infty}p_k x^k$, ($p_0=1$, $p_i=0, i<0$), the polynomials $P(x)$ and $Q(x)$ of degree $m$ and $n$ for the approximation condition (\ref{eq:pade1a}) are given by
\begin{equation}\label{eq:PQ1a}
P(x)=\displaystyle\sum_{i=0}^{m}s_{(m^n, i)}x^i,\quad Q(x)=\displaystyle\sum_{i=0}^{n}s_{((m+1)^i, m^{n-i})}(-x)^i,
\end{equation}
where $s_{\lambda}$ is the Schur function defined by the Jacobi Trudi formula
\begin{equation}\label{eq:JTa}
s_{(\lambda_1, \ldots, \lambda_l)}:=\det(p_{\lambda_i-i+j})_{i, j=1}^l, 
\end{equation}
and $m^n=(\underbrace{m,m,\ldots,m}_n)$.
\end{prop}
For the proof, see Section 2 of \cite{Yamada09} for example.

\begin{lem}\label{lem:PQ2a}
The polynomials $P(x)$ and $Q(x)$ in Proposition \ref{prop:PQ1a} can be expressed in terms of a single determinant as 
\begin{equation}\label{eq:PQ2a}
P(x)=x^ms_{(m^{n+1})}|_{p_i \rightarrow \sum_{j=0}^{i}x^{-j}p_{i-j}}, \quad  Q(x)=(-x)^n s_{((m+1)^n)}|_{p_i \rightarrow p_i -x^{-1}p_{i-1}}.
\end{equation}
\end{lem}
\prf
Direct computation of the right hand side of eqs.(\ref{eq:PQ2a}).
$\square$

Note that the normalization of the polynomials $P(x)$ and $Q(x)$ in eqs.(\ref{eq:PQ1a}) and (\ref{eq:PQ2a}) are different from the convention $P(0)=1$ in the approximation condition (\ref{eq:pade1a}). However, this
difference does not affect the results in the following Proposition \ref{prop:FGa}, since the common normalization factors of $P(x)$ and $Q(x)$ are cancels in eqs.(\ref{eq:F2a})$-$(\ref{eq:G4a}). 

We apply the general results described above to the function $\psi(x)$ in eq.(\ref{eq:psia})
which can be written as
\begin{equation}\label{eq:Tsuda}
\psi(x)=\sum_{k=0}^{\infty}p_k x^k=\exp\Big(\sum_{k=1}^{\infty}\sum_{s=1}^{N+1}\dfrac{b_s^{k}-a_s^{k}}{k(1-q^k)}x^k\Big).
\end{equation}
We note that this kind of the expression (\ref{eq:Tsuda}) has already appeared in \cite{Tsuda10}.

By the definition (\ref{eq:psia}), it is easy to see the properties for $p_k$ as follows:
\begin{equation}\label{eq:p1}
\begin{array}{c}
T_{a_s}^{-1}(p_i)=p_i-\dfrac{1}q{a_s}p_{i-1},\quad T_{b_s}(p_i)=p_i-b_s p_{i-1},\\
\displaystyle T_{a_s}(p_i)=\sum_{j=0}^{i}a_s^{j}p_{i-j},\quad T_{b_s}^{-1}(p_i)=\sum_{j=0}^{i}(\frac{b_s}{q})^{j}p_{i-j},
\end{array}
\end{equation}
for $s=1,\ldots,N+1$.

\begin{prop}\label{prop:taua}
The polynomials $P(x)$ and $Q(x)$ have the following special values:
\begin{equation}\label{eq:PQva}
\begin{array}l
\ds P\Big(\frac{1}{a_s}\Big)=\Big(\frac{1}{a_s}\Big)^mT_{a_s}(\tau_{m,n+1}),\quad Q\Big(\frac{q}{a_s}\Big)=\Big(-\frac{q}{a_s}\Big)^nT_{a_s}^{-1}(\tau_{m+1,n})\\[5mm]
\ds P\Big(\frac{q}{b_s}\Big)=\Big(\frac{q}{b_s}\Big)^mT_{b_s}^{-1}(\tau_{m,n+1}),\quad Q\Big(\frac{1}{b_s}\Big)=\Big(-\frac{1}{b_s}\Big)^nT_{b_s}(\tau_{m+1,n}),
%\o{P}(a_j)=a_j^mT_{a_j}^{-1}(\o{\tau}_{m,n+1}),\quad \o{Q}(b_j)=(-b_j)^nT_{b_j}^{-1}(\o{\tau}_{m+1,n}),\\
%\o{P}(b_1)=b_1^mT_{b_1}(\o{\tau}_{m,n+1}),\quad \o{Q}(a_1)=(-a_1)^nT_{a_1}(\o{\tau}_{m+1,n}),
\end{array}
\end{equation}
for $s=1,\ldots,N+1$.
Here $\tau_{m,n}$ is defined as
\begin{equation}\label{eq:taua}
\tau_{m,n}=s_{(m^n)}={\rm det} (p_{m-i+j})_{i,j=1}^{n}.
\end{equation}
\end{prop}
\prf
Follows from the properties (\ref{eq:p1}) and the formula (\ref{eq:PQ2a}).
$\square$

\begin{prop}\label{prop:FGa}
The polynomials $F(f,x)$ and $G(g,x)$ are determined by the following special values at $x=\frac{1}{a_i}, \frac{1}{b_i}$:
\begin{equation}\label{eq:F1a}
\dfrac{F(f,\frac{1}{a_i})}{F(f,\frac{1}{b_j})}=\alpha\dfrac{T_{a_i}(\tau_{m,n+1})T_{a_i}^{-1}(\tau_{m+1,n})}
{T_{b_j}^{-1}(\tau_{m,n+1})T_{b_j}(\tau_{m+1,n})} \quad (i, j=1, \ldots, N+1),
\end{equation}
\begin{equation}\label{eq:G1a}
G(g,\frac{1}{a_i})=\beta\dfrac{T_{a_i}(\o{\tau}_{m,n+1})T_{a_i}^{-1}(\tau_{m+1,n})}
{T_{a_1}(\tau_{m,n+1})T_{a_1}^{-1}(\o{\tau}_{m+1,n})} \quad (i=2, \ldots, N+1).
\end{equation}
\begin{equation}\label{eq:G2a}
G(g,\frac{1}{b_i})=\gamma\dfrac{T_{b_i}^{-1}(\tau_{m,n+1})T_{b_i}(\o{\tau}_{m+1,n})}
{T_{b_1}^{-1}(\o{\tau}_{m,n+1})T_{b_1}(\tau_{m+1,n})} \quad (i=2, \ldots, N+1), 
\end{equation}
where
\begin{equation}
\begin{array}l
\alpha=-q^{n-m}\dfrac{a_i}{b_j}\dfrac{B(\frac{1}{a_i})}{A(\frac{1}{b_j})},\quad \beta=-q^n \dfrac{a_i}{a_1}\dfrac{B_1(\frac{1}{a_i})}{(\frac{b_1}{a_1})_1}
,\quad \gamma=-q^m \dfrac{b_i}{b_1}\dfrac{A_1(\frac{1}{b_i})}{(\frac{a_1}{b_1})_1}.
\end{array}
\end{equation}
\end{prop}
\prf
From the first equation of (\ref{eq:Da}), we have
\begin{equation}\label{eq:F2a}
\dfrac{F(f,\frac{1}{a_i})}{F(f,\frac{1}{b_j})}=-\Big(\dfrac{a_i}{b_j}\Big)^{m+n+1}\dfrac{B(\frac{1}{a_i})}{A(\frac{1}{b_j})}\dfrac{P(\frac{1}{a_i})Q(\frac{q}{a_i})}
{P(\frac{q}{b_j})Q(\frac{1}{b_j})} \quad (i, j=1, \ldots, N+1).
\end{equation}
From the second and third equations of (\ref{eq:Da}), we have
\begin{equation}\label{eq:G3a}
\begin{array}l
G(g,\frac{1}{a_i})=-\Big(\dfrac{a_i}{a_1}\Big)^{m+n+1}\dfrac{B_1(\frac{1}{a_i})}{(\frac{b_1}{a_1})_1}\dfrac{\o{P}(\frac{1}{a_i})Q(\frac{q}{a_i})}
{P(\frac{1}{a_1})\o{Q}(\frac{1}{a_1})}, \quad (i=2, \ldots, N+1)
\end{array}
\end{equation}
\begin{equation}\label{eq:G4a}
\begin{array}l
G(g,\frac{1}{b_i})=-\Big(\dfrac{b_i}{b_1}\Big)^{m+n+1}\dfrac{A_1(\frac{1}{b_i})}{(\frac{a_1}{b_1})_1}\dfrac{P(\frac{q}{b_i})\o{Q}(\frac{1}{b_i})}
{\o{P}(\frac{1}{b_1})Q(\frac{1}{b_1})}. \quad (i=2, \ldots, N+1)
\end{array}
\end{equation}
Substituting the special values (\ref{eq:PQva}) into the expressions (\ref{eq:F2a})$-$(\ref{eq:G4a}) respectively, we obtain eqs.(\ref{eq:F1a})$-$(\ref{eq:G2a}).
$\square$

We remark that the function $p_k$ can be written in terms of the $q$-Appell Lauricella function %(i.e., the multivariable $q$- hypergeometric function) 
$\varphi_D^{(l)}$ \cite{GaR04} as follows:
%proposition
\begin{prop}
The function $p_k$ can be explicitly written as
\begin{equation}\label{eq:Appell}
p_k=\frac{b_{N+1}^k\Big(\frac{a_{N+1}}{b_{N+1}}\Big)_k}{(q)_k}\varphi_D^{(N)} \Big(q^{-k},\frac{a_1}{b_1},\ldots,\frac{a_N}{b_N},q^{-k+1}\frac{b_{N+1}}{a_{N+1}};q\frac{b_1}{a_{N+1}},\ldots,q\frac{b_N}{a_{N+1}}\Big),
\end{equation} 
\begin{equation}\label{eq:appell}
\varphi_D^{(l)}(\alpha,\beta_1,\ldots,\beta_l,\gamma;z_1,\ldots,z_l):=\displaystyle\sum_{{m_i} \geq 0}\dfrac{(\alpha)_{|m|}(\beta_1)_{m_1}\ldots(\beta_l)_{m_l}}{(\gamma)_{|m|}(q)_{m_1}\ldots(q)_{m_l}}z_1^{m_1}\ldots z_l^{m_l},
\end{equation} 
where $|m|=m_1+\ldots+m_{l}$.    
\end{prop}
%proof
\prf
By the definition of $\psi(x)$ (\ref{eq:psia}) and the $q$-binomial theorem, we have
\begin{equation}\label{eq:Taylor}
\psi(x) =\sum_{m_i \geq 0}
\dfrac{\Big(\frac{a_1}{b_1}\Big)_{m_1} \ldots \Big(\frac{a_{N+1}}{b_{N+1}}\Big)_{m_{N+1}}}{(q)_{m_1}\ldots (q)_{m_{N+1}}}b_1^{m_1}\ldots b_{N+1}^{m_{N+1}}x^{m_1+\cdots+m_{N+1}}.
\end{equation}
%On the other hand, we have the following expressions 
%\begin{equation}\label{eq:poch}
%\dfrac{\Big(\dfrac{b_{N+1}}{a_{N+1}}\Big)_{|m|-m_1- \ldots -m_N}}{(q)_{|m|-m_1- \ldots -m_N}}=\dfrac{\Big(\dfrac{b_{N+1}}{a_{N+1}}\Big)_{|m|} q^{m_1 +\ldots +m_N}(q^{-|m|})_{m_1 + \ldots +m_N}}{(q)_{|m|}\Big(\dfrac{b_{N+1}}{a_{N+1}}\Big)^{m_1 +\ldots +m_N}\Big(\dfrac{b_{N+1}}{a_{N+1}}q^{-|m|+1}\Big)_{m_1 + \ldots +m_N}}.
%\end{equation}
Note that for $k\geq m_{N+1}$, we have
\begin{equation}\label{eq:poch}
\dfrac{\Big(\frac{a_{N+1}}{b_{N+1}}\Big)_{m_{N+1}}}{(q)_{m_{N+1}}}=
\dfrac{\Big(\frac{a_{N+1}}{b_{N+1}}\Big)_{k} 
(q^{-k})_{k-m_{N+1}}}
{(q)_{k}\Big(q^{-k+1}\frac{b_{N+1}}{a_{N+1}}\Big)_{k-m_{N+1}}}\Big(q\frac{b_{N+1}}{a_{N+1}}\Big)^{k-m_{N+1}}.
\end{equation}
Substituting eq.(\ref{eq:poch}) with $k=|m|+m_{N+1}$ into eq.(\ref{eq:Taylor}), we obtain eq.(\ref{eq:Appell}).
$\square$

In \cite{Sakai05-2}, a hypergeometric solution of the $q$-Garnier system is given in terms of the $q$-Appell Lauricella function $\varphi_D^{(N)}$ (\ref{eq:appell}). Our result corresponds to its determinantal generalization in terminating case.
For the differential Garnier system, a more general determinant formula applicable also to the transcendental solutions is derived by 
%the construction of a Schlesinger transformation 
applying the (Hermite-)Pad\'e approximation \cite{Mano12, MT14}.

\section{The Pad\'e problem on $q$-grid}\label{sec:padeq}

In this section, we study certain Pad\'e interpolation  problem and solve it explicitly. As a result, we obtain some special solutions of the $q$-Garnier system given in terms of the generalized $q$-hypergeometric function. 
%We remark that similar computations in subsection \ref{subsec:LGq} and \ref{subsec:solq} have been already done in \cite{Ikawa13, Nagao15-1,NTY13}. 
\subsection{Lax pair and the $q$-Garnier system}\label{subsec:LGq}

In this subsection, starting the Pad\'e interpolation problem (\ref{eq:pade1q}), we derive linear difference relations (\ref{eq:L2L3q})
% and (\ref{eq:L1q}) 
and nonlinear relations (\ref{eq:gdq})$-$(\ref{eq:fu2q}).

%Fix a positive integer $N$ and a complex parameter $q$ ($0 < |q| <1$).
For complex parameters $a_1, \ldots, a_{N}, b_1, \ldots, b_{N}, c \in \C^{\times}$, we put
\begin{equation}\label{eq:psiq}
\psi(x):=c^{\log_q x}\prod_{i=1}^{N}\dfrac{(a_i x, b_i)_\infty}{(a_i, b_i x)_\infty}.
%, \quad \psi_s:=\psi(q^s)=c^s\prod_{i=1}^{N}\dfrac{(b_i)_s}{(a_i)_s}.
\end{equation}
%Here and in what follows, we use the standard $q$-Pochhammer symbols defined as
%\begin{equation}
%(z)_\infty=\prod_{i=0}^\infty (1-q^i z), \quad
%(z)_s=\frac{(z)_\infty}{(zq^s)_\infty},  \quad
%(z_1,z_2, \ldots)_s=(z_1)_s(z_2)_s \ldots.
%\end{equation}

Define polynomials $P(x)$ and $Q(x)$ of degree $m$ and $n$ $\in \Z_{\ge 0}$ by the following Pad\'e interpolation condition:
\begin{equation}\label{eq:pade1q}
\psi(x_s)=\dfrac{P(x_s)}{Q(x_s)} \quad (x_s=q^s, s=0, 1, \ldots m+n)
\end{equation}
%This is the Pad\'e problem which is the bases of our study. 
The common normalizations of the polynomials $P(x)$ and $Q(x)$ are fixed as $P(0)=1$ tentatively.  As in Section {\ref{subsec:LGa},
%we introduce a discrete transformation $T$ of the parameters and study the contiguity relations between the original and shifted objects.
the shift $T$ is given by eq.(\ref{eq:Ta}). 
%\begin{equation}\label{eq:Tq}
%T:=T_{a_1}T_{b_1},\quad T_a: a \to qa.
%\end{equation}
%For any object $X$, the corresponding shifts are denoted as $\o{X}:=T(X)$, $\u{X}:=T^{-1}(X)$. 
%For example, the symbols $\o{P}(x)$, $\o{Q}(x)$ represent the polynomials defined by 
%\begin{equation}\label{eq:pade2q}
%\o{\psi}(x_s)=\dfrac{\o{P}(x_s)}{\o{Q}(x_s)} \quad (x_s=q^s, s=0, 1, \ldots m+n).
%\end{equation}
%We call the operators $T$ as the "time evolution", since it will play the role of time evolution of the $q$-Garnier system.

%Let us consider two more linear three term relations: $L_2(x)$ between $y(x), y(qx), \o{y}(x)$ and $L_3(x)$ between $y(x), \o{y}(x), \o{y}(x/q)$ satisfied by the functions $y(x)=P(x)$ and $y(x)=\psi(x)Q(x)$. 

\begin{prop}\label{prop:L2L3q}
For $y(x)=P(x)$ and $y(x)=\psi(x)Q(x)$, we have the following linear relations: 
\begin{equation}\label{eq:L2L3q} 
\begin{array}l
\ds L_2(x)=(g_0)_1F(f,x)\o{y}(x)-(\frac{x}{q^{m+n}})_1 A_1(x)y(qx)+(b_1x)_1 G(g,x)y(x)=0,\\[3mm]
\ds L_3(x)=(\frac{g_0}{c})_1 F(\o{f},\frac{x}{q})y(x)+\frac{1}{c}(a_1x)_1G(g,\frac{x}{q})\o{y}(x)-(x)_1 B_1(\frac{x}{q})\o{y}(\frac{x}{q})=0,
\end{array}
\end{equation}
where
\begin{equation}
\begin{array}l
\ds A(x)=\prod_{j=1}^{N}(a_j x)_1,\quad \ds B(x)=\prod_{j=1}^{N}(b_j x)_1,\quad \ds F(f,x)=1+\sum_{j=1}^N f_j x^j,\\
A_1 (x)=\dfrac{A(x)}{(a_1 x)_1}, \quad B_1 (x)=\dfrac{B(x)}{(b_1 x)_1}, \quad \ds G(g,x)=\sum_{j=0}^{N-1} g_j x^{j}.
\end{array}
\end{equation}
Here $\o{f}_i=T(f_i)$, and $f_1, \ldots, f_N, g_0, \ldots, g_{N-1}\in\P^{1}$ are some constants depending on parameters $a_i$, $b_i$, $c$, $m$, $n$. 
\end{prop}

\prf
%The proof is similar to the proof of theorem 2.1 of \cite{Ikawa13}. Similarly to proposition \ref{prop:L1q}, 
The method of the proof is the same as that of Proposition \ref{prop:L2L3a}. By the definition of the linear relations $L_2(x)=0$ and $L_3(x)=0$, they can be written as the expression (\ref{eq:L2L3matrixa}).
%{\small
%\begin{align}\label{eq:L2L3matrixq}
%L_2(x)=
%\begin{vmatrix}
%y(x) & y(qx) & \o{y}(x) \\
%P(x) & P(qx) & \o{P}(x)\\
%\psi(x)Q(x) & \psi(qx)Q(qx) & \o{\psi}(x)\o{Q}(x)
%\end{vmatrix}=0, \quad
%L_3(x)=
%\begin{vmatrix} 
%y(x) & \o{y}(x) & \o{y}(x/q) \\
%P(x) & \o{P}(x) & \o{P}(x/q)\\
%\psi(x)Q(x) & \o{\psi}(x)\o{Q}(x) & \o{\psi}(x/q)\o{Q}(x/q)
%\end{vmatrix}=0.
%\end{align}
%}
Define Casorati determinants $D_1(x)$, $D_2 (x)$ and $D_3(x)$ by eq.(\ref{eq:Ddefa}).
%\begin{equation}\label{eq:D2D3defq}
%\begin{array}{l}
%D_2(x)=\det[{\bf y}(x),{\o{\bf y}}(x)],\quad D_3(x)=\det[{\bf y}(qx),\o{{\bf y}}(x)].
%\end{array}
%\end{equation}
Then, the linear relations $L_2(x)=0$ and $L_3(x)=0$ can take the forms (\ref{eq:L2L3Da}).
%\begin{equation}\label{eq:L2L3Dq}
%\begin{array}{l}
%L_2(x): D_1(x) \o{y}(x)-D_2(x)y(qx)+D_3(x)y(x)=0,
%\\
%L_3(x): \o{D}_1(\frac{x}{q})y(x)+D_3(\frac{x}{q}) \o{y}(x)-D_2(x)\o{y}(\frac{x}{q})=0.
%\end{array}
%\end{equation}
The determinants (\ref{eq:Ddefa}) can be computed by the condition (\ref{eq:pade1q}) and the relations
\begin{equation}
\dfrac{\psi(qx)}{\psi(x)}=c\dfrac{B(x)}{A(x)},
\quad \dfrac{\o{\psi}(x)}{\psi(x)}=\dfrac{(a_1 ,b_1 x)_1}{(a_1 x, b_1)_1}.
\end{equation}
The results are given as 
\begin{equation}\label{eq:Dq}
\begin{array}{l}
D_1(x)=\dfrac{\psi(x)}{A(x)}\left\{cB(x)P(x)Q(qx)-A(x)P(qx)Q(x)\right\}\\
\phantom{D_1(x)}
=:w_0 \dfrac{\psi(x)\prod_{i=0}^{m+n-1}(\frac{x}{q^i})_1}{A(x)} F(f,x),\\
D_2(x)=\dfrac{\psi(x)}{(a_1 x, b_1)_1}\left\{(a_1 ,b_1 x)_1P(x)\o{Q}(x)-(a_1 x,b_1)_1\o{P}(x)Q(x)\right\}\\[5mm]
\phantom{D_2(x)}=:w_1\dfrac{\psi(x)\prod_{i=0}^{m+n}(\frac{x}{q^i})_1}{(a_1 x, b_1)_1},\\
D_3(x)=\dfrac{\psi(x)}{A(x)(b_1)_1}\left\{(a_1 ,b_1 x)_1A_1(x)P(qx)\o{Q}(x)-c(b_1)_1B(x)\o{P}(x)Q(qx)\right\}\\[5mm]\phantom{D_3(x)}=:w_1\dfrac{\psi(x)\prod_{i=0}^{m+n-1}(\frac{x}{q^i})_1}{A(x)(b_1)_1}(b_1 x)_1G(g,x),
\end{array}
\end{equation}
with some constants $w_0$ and $w_1$ depending on parameters $a_i$, $b_i$, $c$, $m$ and $n$. 
%Substituting eqs.(\ref{eq:Dq}) into eq.(\ref{eq:L2L3Da}), 
%and using the suitable transformation of $y(x)$ in $L_2$ i.e. $y(x)\mapsto Gy(x)$, $G/\o{G}=1/c$, $w=r/c^2$, 
%we obtain eq.(\ref{eq:L2L3q}), where
The constants $w_0$ and $w_1$ are fixed as $w_0=(g_0)_1$ and $w_1=(g_0/c)_1$ by the condition that eq.(\ref{eq:L2L3Da}) has a solution such as $y(0)=P(0)=1$.
$\square$

These relations (\ref{eq:L2L3q}) are regarded as the Lax equations $L_2(x)=0$ and $L_3(x)=0$  for the $q$-Garnier system (cf. eq.(\ref{eq:L2L3})).

%We will derive the birational evolution equations among $f_i, g_i, \o{f}_i, \u{g}_i$ from the compatibility of $L_2, L_3$ (\ref{eq:L2L3}).
\begin{prop}\label{prop:Geqq}
The constants $f_1, \ldots ,f_N$ and $g_0, \ldots ,g_{N-1}$ satisfy the following relations:
%The compatibility of the relations $L_2$ and $L_3$ (\ref{eq:L2L3q}) gives the following condition:
\begin{equation}\label{eq:gdq}
(qx,\dfrac{x}{q^{m+n}})_1 A_1 (x)B_1 (x) -\dfrac{1}{c}(a_1 x, b_1 x)_1 G(g,x)G(\u{g},x)=0
\quad {\rm for} \quad F(f,x)=0,
\end{equation}
\begin{equation}\label{eq:fu1q}
(qx,\dfrac{x}{q^{m+n}})_1 A_1 (x)B_1 (x)-(g_0,\dfrac{g_0}{c})_1F(f,x)F(\o{f},x)=0
\quad {\rm for} \quad G(g,x)=0,
\end{equation}
\begin{equation}\label{eq:fu2q}
(g_0,\dfrac{g_0}{c})_1f_N \o{f}_N =\Big(\frac{q a_1}{c}g_{N-1}-\frac{\prod_{i=2}^{N}(-b_i)}{q^{m-1}}\Big)\Big(b_1 g_{N-1}-\frac{\prod_{i=2}^{N}(-a_i)}{q^n}\Big),
\end{equation}
%where $c$ (resp. $q^m$) is one of the exponents of the linear equation $L_1(x)$ at $x=0$ (resp. $x=\infty$). 
\end{prop}
 
\prf 
Similar to the proof of Proposition \ref{prop:Geq}.
%Eliminating $y(x), y(qx)$ from $L_2(x)=\u{L}_3(qx)=0$ for $F(f,x)=0$, we obtain the relation (\ref{eq:gdq}). Similarly, eliminating $y(qx), \u{y}(x)$ from $L_2(x)=L_3(qx)=0$ for $G(g,x)=0$, we obtain the relation (\ref{eq:fu1q}). Considering the highest coefficients of $L_2(x)$ and $L_3(x)$ for $y(x)=P(x)$, we obtain the relation (\ref{eq:fu2q}). 
$\square$

These relations (\ref{eq:gdq})$-$(\ref{eq:fu2q}) are also regarded as the $q$-Garnier system (cf. eqs.(\ref{eq:gd})$-$(\ref{eq:fu2})).

\subsection{Special solutions}\label{subsec:solq}

We derive the explicit forms (\ref{eq:F1q})$-$(\ref{eq:G2q}) of variables $\{f_i, g_i\}$ appearing in the Casorati determinants $D_1(x)$ and $D_3(x)$ (\ref{eq:Dq}). They are interpreted as the special solutions for the $q$-Garnier system (\ref{eq:gdq})$-$(\ref{eq:fu2q}) due to Proposition \ref{prop:Geqq}.

%We will derive formulae (\ref{eq:qJacobi}) which are convenient for computing the special solutions $f$ and $g$. The Jacobi formulae (\ref{eq:Jacobi}) are essentially presented in \cite{Jacobi}. 

\begin{prop}\label{prop:Jacobi}(\cite{Jacobi} also \cite{Nagao15-1})
For a given sequence $\psi_s$, the polynomials $P(x)$ and $Q(x)$ of degree $m$ and $n$ for an interpolation problem 
\begin{equation} \label{eq:pade3q}
\psi_s=\frac{P (x_s)}{Q (x_s)} \quad (s=0, 1, \dots, m+n)
\end{equation}
are given by the following determinant expressions:
\begin{equation}\label{eq:Jacobi}
P(x)=F(x)\det\Big[\sum^{m+n} _{s=0}u_s \dfrac{x_s^{i+j}}{x-x_s }\Big]^n _{i,j=0}, \quad Q(x)=\det\Big[\sum^{m+n} _{s=0}u_s x_s^{i+j}(x-x_s )\Big]^{n-1} _{i,j=0},
\end{equation}
where $u_s=\frac{\psi_s}{F^{\prime}(x_s)}$ and $F(x)=\prod_{i=0}^{m+n}(x-x_i )$.
\end{prop}

\begin{lem}\label{lem:qJacpbi}(\cite{Nagao15-1})
In the $q$-grid case of problem (\ref{eq:pade3q}) (i.e., $x_s=q^s$), the formulae (\ref{eq:Jacobi}) take the following form:
\begin{equation}\label{eq:qJacobi}
\begin{array}{l}
P(x)=\dfrac{F(x)}{(q)_{m+n}^{n+1}}\det\Big[\displaystyle\sum^{m+n} _{s=0}\psi_s\dfrac{(q^{-(m+n)})_s }{(q)_s}\dfrac{q^{s(i+j+1)}}{x-q^s }\Big]^n _{i,j=0},
\\
Q(x)=\dfrac{1}{(q)_{m+n}^{n}}\det\Big[\displaystyle\sum^{m+n} _{s=0}\psi_s\dfrac{(q^{-(m+n)})_s }{(q)_s }q^{s(i+j+1)}(x-q^s)\Big]^{n-1} _{i,j=0}.
\end{array}
\end{equation}
\end{lem}

\prf
In the derivation of (\ref{eq:qJacobi}), we have used the relations
\begin{align}\label{eq:Fprime}
F^{\prime}(x_s )
%=&(x_s -x_0 )\dots(x_s -x_{s-1})(x_s -x_{s+1})\dots(x_s -x_{m+n})\nonumber \\
%=&(-1)^s q^{(s-1)s/2}(q;q)_s q^{s(m+n-s)}(q;q)_{m+n-s}\nonumber\\
=&\frac{(q)_s (q)_{m+n}}{q^{s}(q^{-(m+n)})_s}.  
\end{align}
Substituting the value of $F^{\prime}(x_s )$ (\ref{eq:Fprime}) into the formulae (\ref{eq:Jacobi}), one obtains the determinant formulae (\ref{eq:qJacobi}).
$\square$

The normalization of the polynomials $P(x)$ and $Q(x)$ in eqs.(\ref{eq:Jacobi}) and (\ref{eq:qJacobi}) are different from the convention $P(0)=1$ in the interpolation condition (\ref{eq:pade1q}). As in Section {\ref{subsec:sola}, this
difference does not affect the results in the following Proposition \ref{prop:FGq}.
%since the common normalization factors of $P(x)$ and $Q(x)$ are cancels in eq.(\ref{eq:F2q})$-$(\ref{eq:G4q}). 
\begin{prop}\label{prop:tauq}
The polynomials $P(x)$ and $Q(x)$ defined in Section \ref{subsec:LGq} have the following special values:
\begin{equation}\label{eq:PQvq}
\begin{array}l
P(\frac{1}{a_s})=\dfrac{(a_s)_{m+n+1}}{a_s^m (a_s)_1^{n+1}(q)_{m+n}^{n+1}}T_{a_s}(\tau_{m,n}),\quad Q(\frac{q}{a_s})=\dfrac{q^n (\frac{a_s}{q})_1^n}{a_s^n (q)_{m+n}^n}T_{a_s}^{-1}(\tau_{m+1,n-1})\\[5mm]
P(\frac{q}{b_s})=\dfrac{q^m (\frac{b_s}{q})_{m+n+1}}{b_s^m (\frac{b_s}{q})_1^{n+1}(q)_{m+n}^{n+1}}T_{b_s}^{-1}(\tau_{m,n}),\quad Q(\frac{1}{b_s})=\dfrac{(b_s)_1^n}{b_s^n (q)_{m+n}^n}T_{b_s}(\tau_{m+1,n-1}),
%\o{P}(a_j)=a_j^mT_{a_j}^{-1}(\o{\tau}_{m,n+1}),\quad \o{Q}(b_j)=(-b_j)^nT_{b_j}^{-1}(\o{\tau}_{m+1,n}),\\
%\o{P}(b_1)=b_1^mT_{b_1}(\o{\tau}_{m,n+1}),\quad \o{Q}(a_1)=(-a_1)^nT_{a_1}(\o{\tau}_{m+1,n}),
\end{array}
\end{equation}
for $s=1,\ldots,N$.
Here $\tau_{m,n}$ is defined as
\begin{equation}\label{eq:tauq}
\tau_{m,n}=\det\Big[{}_{N+1}\varphi_N \Big(\substack{\displaystyle{b_1, \ldots, b_N ,q^{-(m+n)}}\\[3mm]{\displaystyle{a_1 ,\ldots, a_N}}},cq^{i+j+1}\Big)\Big]^n _{i,j=0},
\end{equation}
and the $q$-HGF (the $q$-hypergeometric functions \cite{GaR04}) is defined by
\begin{equation}\label{eq:qHGF}
\begin{array}{l}
{}_k\varphi_l\left(
\begin{array}{ccc}
\alpha_1,&\dots,&\alpha_k\\[0mm]
\beta_1,&\dots,&\beta_l
\end{array}
,x
\right)
:=\displaystyle\sum_{s=0}^{\infty}\dfrac{(\alpha_1,\ldots,\alpha_k)_s}{(\beta_1,\ldots,\beta_l,q)_s}\left[(-1)^sq^{\left(\substack{s\\2}\right)}\right]^{1+l-k}x^s,
\end{array}
\end{equation}
with $\left(\substack{s\\2}\right)=\frac{s(s-1)}{2}$.

\end{prop}
\prf
Follows from the formula (\ref{eq:qJacobi}) and the sequence $\psi_s=\psi(q^s)=c^s\prod_{i=1}^N\frac{(b_i)_s}{(a_i)_s}$.
$\square$

\begin{prop}\label{prop:FGq}
The polynomials $F(f,x)$ and $G(g,x)$ are determined by the following special values at $x=\frac{1}{a_i}, \frac{1}{b_i}$:
\begin{equation}\label{eq:F1q}
\dfrac{F(f,\frac{1}{a_i})}{F(f,\frac{1}{b_j})}=\alpha\dfrac{T_{a_i}(\tau_{m,n})T_{a_i}^{-1}(\tau_{m+1,n-1})}
{T_{b_j}^{-1}(\tau_{m,n})T_{b_j}(\tau_{m+1,n-1})} \quad (i, j=1, \ldots, N),
\end{equation}
\begin{equation}\label{eq:G1q}
G(g,\frac{1}{a_i})=\beta\dfrac{T_{a_i}(\o{\tau}_{m,n})T_{a_i}^{-1}(\tau_{m+1,n-1})}
{T_{a_1}(\tau_{m,n})T_{a_1}^{-1}(\o{\tau}_{m+1,n-1})} \quad (i=2, \ldots, N),
\end{equation}
\begin{equation}\label{eq:G2q}
G(g,\frac{1}{b_i})=\gamma\dfrac{T_{b_i}^{-1}(\tau_{m,n})T_{b_i}(\o{\tau}_{m+1,n-1})}
{T_{b_1}^{-1}(\o{\tau}_{m,n})T_{b_1}(\tau_{m+1,n-1})} \quad (i=2, \ldots, N), 
\end{equation}
\end{prop}
where
\begin{equation}
\begin{array}l
\alpha=-cq^{n-m}\dfrac{(a_i q^{m+n})_1 (\frac{b_j}{q})_1^n (\frac{a_i}{q})_1^n}{(a_i)_1^{n+1}(b_j)_1^n}\dfrac{B(\frac{1}{a_i})}{A(\frac{1}{b_j})},\\[5mm]
\beta=c\dfrac{(b_1 , a_i q^{m+n})_1 (\frac{a_i}{q})_1^n B_1 (\frac{1}{a_i})}{a_1 q^m (\frac{b_1}{a_1})_1 (a_i)_1^{n+1}},\quad \gamma=\dfrac{(a_1)_1 (b_i)_1^n A_1(\frac{1}{b_i})}{b_1 q^n (\frac{a_1}{b_1})_1 (\frac{b_i}{q})_1^n}. 
\end{array}
\end{equation}

\prf
Taking the ratio $\frac{D_1 (\frac{1}{a_i})}{D_1 (\frac{1}{b_j})}$ (\ref{eq:Dq}), we have
\begin{equation}\label{eq:F2q}
\ds \dfrac{F(f,\frac{1}{a_i})}{F(f,\frac{1}{b_j})}=-c\prod_{s=0}^{m+n-1}\dfrac{(\frac{1}{b_j q^s})_1}{(\frac{1}{a_i q^s})_1}\dfrac{B(\frac{1}{a_i})}{A(\frac{1}{b_j})}\dfrac{P(\frac{1}{a_i})Q(\frac{q}{a_i})}
{P(\frac{q}{b_j})Q(\frac{1}{b_j})} \quad (i, j=1, \ldots, N).
\end{equation}
Taking the ratio $\frac{D_3 (\frac{1}{a_i})}{D_2 (\frac{1}{a_1})}$ (\ref{eq:Dq}), we have
\begin{equation}\label{eq:G3q}
\begin{array}l
\ds G(g,\frac{1}{a_i})=-\dfrac{c(b_1)_1 B_1(\frac{1}{a_i})}{(a_1 ,\frac{b_1}{a_1})_1}\dfrac{\prod_{s=0}^{m+n}(\frac{1}{a_1 q^s})_1}{\prod_{s=0}^{m+n-1}(\frac{1}{a_i q^s})_1}\dfrac{\o{P}(\frac{1}{a_i})Q(\frac{q}{a_i})}
{P(\frac{1}{a_1})\o{Q}(\frac{1}{a_1})} \quad (i=2, \ldots, N)
\end{array}
\end{equation}
Taking the ratio $\frac{D_3 (\frac{1}{a_i})}{D_2 (\frac{1}{a_1})}$ (\ref{eq:Dq}), we have
\begin{equation}\label{eq:G4q}
\begin{array}l
\ds G(g,\frac{1}{b_i})=-\dfrac{\prod_{s=0}^{m+n}(\frac{1}{b_1 q^s})_1}{\prod_{s=0}^{m+n-1}(\frac{1}{b_i q^s})_1}\dfrac{(a_1)_1A_1(\frac{1}{b_i})}{(\frac{a_1}{b_1},b_1)_1}\dfrac{P(\frac{q}{b_i})\o{Q}(\frac{1}{b_i})}
{\o{P}(\frac{1}{b_1})Q(\frac{1}{b_1})} \quad (i=2, \ldots, N).
\end{array}
\end{equation}
Substituting the special values (\ref{eq:PQvq}) into the expressions (\ref{eq:F2q})$-$(\ref{eq:G4q}) respectively, we obtain the values (\ref{eq:F1q})$-$(\ref{eq:G2q}).
$\square$

\begin{rem}{\rm {\bf On the relation between special solutions in Sections \ref{sec:padea} and \ref{sec:padeq}}} There exists the following relation between the $q$-Appell Lauricella function $\varphi_D^{(l)}$ and the $q$-hypergeometric function ${}_{l+1}\varphi_l$(see \cite{Andrews72, Andrews75, Kajihara04} for example):
\begin{equation}\label{eq:Andorews}
\begin{array}{l}
{}_{l+1}\varphi_l\left(
\begin{array}{ccc}
a,&b_1\dots,&b_l\\[0mm]
&c_1,\dots,&c_l
\end{array}
,u
\right)=\dfrac{(au)_{\infty}}{(u)_{\infty}}\displaystyle\prod_{k=1}^l \dfrac{(b_k)_{\infty}}{(c_k)_{\infty}}\varphi_D^{(l)}(u,\frac{c_1}{b_1},\ldots,\frac{c_l}{b_l},au;b_1,\ldots,b_l).
\end{array}
\end{equation}
%Concerning to the results of this paper, this formula is applicable to the terminating cases (\ref{eq:tauq}) and (\ref{eq:Appell}) appearing in this paper.
%The formula above can not be applied to the two terminating functions ${}_{N+1}\varphi_N$ in (\ref{eq:tauq}) and $\varphi_D^N$ in (\ref{eq:Appell}) and the relation with the terminating functions can not be confirmed.

In view of this, one can expect that the special solutions in Sections \ref{sec:padea} and \ref{sec:padeq} may have some relations. The study of the relations is an interesting future problem.
%The relations are an interesting problem for the further study. 
%coincide with each other, however this is not the case. Namely if we put $a_{N+1}=\frac{1}{q^{m+n}}$,  $b_{N+1}=q$ in Section \ref{sec:padea} and $c=q^{m+n+1}$ in Section \ref{sec:padeq}, then these special solutions are different under the common specialization of parameters.
%Then both solutions correspond to common parameters but they are different. 
\end{rem}

In \cite{Suzuki15}, some special solution of the higher order $q$-Painlev\'e system is given in terms of the $q$-hypergeometric function ${}_{N+1}\varphi_N$. Our results suggest the relation between the system in \cite{Suzuki15} and $q$-Garnier system. In fact, it turns out that these two are equivalent as will be shown in \cite{NYyokoku}.

\section*{Acknowledgment}

The authors are grateful to Professors Tetsu Masuda, Masatoshi Noumi, Hidetaka Sakai, Takao Suzuki, Takashi Takebe, Teruhisa Tsuda and the referee for stimulating discussions and/or comments. 
This work was partially supported by JSPS KAKENHI (26287018).

\hspace{70mm}nuna adreso:

\hspace{70mm}Hidehito Nagao

\hspace{70mm}Department of Arts and Science

\hspace{70mm}National Institute of Technology

\hspace{70mm}Akashi College

\hspace{70mm}Uozumi, Akashi 674-8501

\hspace{70mm}Japan

\hspace{70mm}nagao@akashi.ac.jp

\hspace{70mm}Yasuhiko Yamada

\hspace{70mm}Department of Mathematics

\hspace{70mm}Graduate School of Science

\hspace{70mm}Kobe University

\hspace{70mm}Rokko, Kobe 657-8501

\hspace{70mm}Japan

\hspace{70mm}yamaday@math.kobe-u.ac.jp

\end{document}